\documentclass[%
 reprint,
 amsmath,amssymb,
 aps,
]{revtex4-2}

\usepackage{graphicx}
\usepackage{dcolumn}
\usepackage{bm}

\begin{document}

\preprint{APS/123-QED}

\title{Accurate computation of the electron-phonon interaction contribution to 
the total energy}

\author{Shilpa Paul}%
\affiliation{%
	Department of Metallurgical Engineering and Materials Science, Indian Institute of Technology Bombay, Mumbai 400076, India
}%

\author{Mogadalai Pandurangan Gururajan}%
\affiliation{%
	Department of Metallurgical Engineering and Materials Science, Indian Institute of Technology Bombay, Mumbai 400076, India
}%

\author{Amrita Bhattacharya}%
\affiliation{%
	Department of Metallurgical Engineering and Materials Science, Indian Institute of Technology Bombay, Mumbai 400076, India
}%
\author{Tiramkudlu R. S. Prasanna}%
\email{prasanna@iitb.ac.in}
\affiliation{%
	Department of Metallurgical Engineering and Materials Science, Indian Institute of Technology Bombay, Mumbai 400076, India
}%

\date{\today}

\begin{abstract}
    The standard Hamiltonian of a coupled electron-phonon system is based on 
    second-order perturbation theory. The EPI contribution in the standard 
    Hamiltonian consists of two terms, the EPI contribution to the 
    band-structure energy and the partial-Fan-Migdal (FM)-occupied 
    contribution. Within the non-adiabatic approximation, we derive a new 
    expression for the partial-FM-occ contribution and show that it has the 
    structure of a higher-order term, and not a second-order term. Along 
    similar lines, we derive new expressions for the computation of the 
    partial-FM-occ term. The new expressions for the partial-FM-occ term must 
    be preferred over the standard expressions, in theoretical and 
    computational studies, because they incorporate the complete physics 
    underlying this term. Unlike the EPI contribution to individual 
    eigenstates, the EPI contribution to the total energy must be computed in 
    the non-adiabatic approximation for all materials, Infra-red (IR) active 
    and IR-inactive. We report the computation of the standard Hamiltonian, for 
    the first time, for Carbon polymorphs (diamond and hexagonal lonsdaleite) 
    by including the EPI contribution to the total energy. This is the most 
    accurate \textit{ab initio} total energy reported till date. The present 
    work also opens the way to  compute the \textit{ab initio} free-energy more 
    accurately at finite temperatures by including the EPI contribution.
     
\end{abstract}

\maketitle


\section{Introduction}
	The total energy is of fundamental importance in \textit{ab initio} studies and is the basis of the density functional theory (DFT) 
	\cite{Kohn1964, Kohn1965}. For crystalline solids, the lattice periodicity is incorporated through the periodic potential, U(\textbf{r}) \cite{Ashcrof1976, Simon2013oxford, Martin2020electronic, Grosso2013solid}. The DFT total energy is minimized for the primitive unit cell and subsequently,  the structural parameters of crystalline materials are obtained. All electronic and optical properties are calculated using the optimized primitive unit cell.
    
	In structural studies on the stable and metastable structures, the \textit{ab initio} total energy is of fundamental importance as it establishes the stable structure and also gives the relative energy differences between polymorphs. There is also great interest at present in the search for new crystal structures, especially, low-lying metastable structures with different properties, of various materials. These studies involve a	search for new metastable structures using a variety of techniques 	\cite{Oganov2019structure, Catlow2020structure, Cheng2022crystal, 	Hong2020training, Yamashita2018crystal}. The \textit{ab initio} total energy is also of fundamental importance in the search for new metastable 	structures.
	
	DFT studies \cite{Kohn1964, Kohn1965} have been successful in determining 
	the stable structure and energy differences between polymorphs for many 
	materials. However, despite this success, there are some materials where there are discrepancies between the predictions of DFT studies and experimental results. For this reason, additional contribution to the \textit{ab initio} total energy are included. The following contributions were previously included i) van der Waals (vdW) energy (computed as part of DFT computation) \cite{Kawanishi2016, Scalise2019, Ramakers, Cazorla2019}) and  ii) zero-point vibrational energy (ZPVE) (computed separately from Density functional perturbation theory (DFPT) or Finite difference method (FDM)	\cite{Kawanishi2016, Scalise2019, Ramakers, Cazorla2019, Van2007, Bhat2018}.

	Clearly, it is essential to determine the \textit{ab initio} total energy 
	as accurately as possible by including all possible contributions. Towards 
	this end, we have recently proposed the inclusion of electron-phonon 
	interaction (EPI) contribution to the total energy \cite{Varma2022}. 
	Subsequently, referring to our work, Allen has derived the expression for 
	the EPI contribution to the total energy using second-order perturbation 
	theory \cite{Allen2020, Allen2022erratum}. This equation, as such, cannot 
	be computed using available software. Hence, in earlier studies, we have 
	approximated Allen's equation \cite{Allen2022erratum} and computed the EPI 
	contribution to the total energy for C, Si, SiC and BN polymorphs 
	\cite{Varma2022, Paul2022}. Our results show that, to determine total 
	energy differences, the EPI contribution is more important than the vdW and 
	ZPVE contributions because it is more sensitive to crystal structure and 
	hence, must be included in all such studies \cite{Varma2022, Paul2022}.
	
	However, for accurate total energy studies, it is necessary to compute the 
	complete Allen's equation (Eq. 4 of Ref. \cite{Allen2022erratum}) for the 
	EPI contribution. In this work, we have recast Allen's equation, for 
	semiconductors and 	insulators, so that it can be computed using existing 
	software. Thus, for the first time, we compute the standard 
	Hamiltonian of a coupled-electron phonon system. This gives the most 
	accurate \textit{ab initio} total energy till date. This 
	result is of great importance for the structural studies discussed above 
	\cite{Oganov2019structure, Catlow2020structure, Cheng2022crystal, 
	Hong2020training, Yamashita2018crystal, Kawanishi2016, Scalise2019, 
	Ramakers, Cazorla2019, Van2007, Bhat2018}.

	It is important to place the recent EPI studies \cite{Varma2022, 
		Allen2022erratum} in perspective. The standard Hamiltonian 
	\cite{Allen-chakraborty1978theory, Giustino2017, Allen2022erratum} for a 
	coupled electron-phonon system is given by \cite{Allen2022erratum} 
	\begin{equation}
		H_{\mathrm{eff}}= \sum_k \epsilon_k(V) c_k^{\dagger} c_k+\sum_Q \hbar 
		\omega_Q\left(a_Q^{\dagger} a_Q+1 / 2\right)+V_{\mathrm{ep}} 
		\label{Std-Hamiltonian-EPI}
	\end{equation}
	\begin{equation}
		\begin{aligned}
			V_{\mathrm{ep}}= &\sum_{k Q} V^{(1)}(k Q) c_{k+Q}^{\dagger} 
			c_k\left(a_Q+a_{-Q}^{\dagger}\right) \\
			&+\sum_{k Q Q^{\prime}} V^{(2)}\left(k Q Q^{\prime}\right) 
			c_{k+Q+Q^{\prime}}^{\dagger} c_k\\ &\mathrm{x} 
			\left(a_Q+a_{-Q}^{\dagger}\right)\left(a_{Q^{\prime}}+a_{-Q^{\prime}}^{\dagger}\right)
			+ \ldots
			\label{Vep}
		\end{aligned}
	\end{equation}
	Eq. \ref{Std-Hamiltonian-EPI} gives the \textit{total energy} of a coupled electron-phonon system. The first two terms give the DFT total energy and the ZPVE contribution respectively \cite{Allen2022erratum,Giustino2022ab}. (The computation of the ZPVE term includes the EPI contribution to the phonon energy, explicitly using DPFT and implicitly using Finite Difference and other methods \cite{Allen2020, Allen2022EPIBS, Heid201312, Giustino2017, Varma2022}.) Crucially, the $V_{ep}$ term, which gives the EPI contribution to the total energy, has never been computed till date because the	emphasis has been to compute the EPI contributions to the individual eigenstates and not to the whole crystal. Even the expression for the $V_{ep}$ term has only recently been derived by Allen \cite{Allen2022erratum} referring to our earlier work \cite{Varma2022}. 

	In this work, we show that Allen's equation for the $V_{ep}$ term 
	\cite{Allen2022erratum}  can be recast so that it contains two 
	contributions, the EPI contribution to the band-structure energy term and a 
	partial-Fan-Migdal-occupied (partial-FM-occ) term. We show that, unlike for 
	the EPI contribution to individual eigenstates, the EPI contribution to the 
	total energy must be computed in the non-adiabatic approximation for all 
	materials,  irrespective of whether they are Infra-red (IR)-active or 
	IR-inactive. We derive a new expression for the partial-FM-occ term in the 
	non-adiabatic approximation that shows that the partial-FM-occ term has the 
	structure of a higher-order term and not a second-order term. Along similar 
	lines, we derive new expressions for the computations of the partial-FM-occ 
	terms that include a smearing parameter. The new expressions must be 
	preferred over the standard expressions in theoretical and computational 
	studies because they incorporate the complete physics of the partial-FM-occ 
	term. 

    In this paper, \textit{for the first time, we report the computation of the standard Hamiltonian for a coupled electron-phonon system.} It is evident that the standard Hamiltonian gives the most accurate \textit{ab initio} total energy.
    
	Further, since the EPI contribution at finite temperatures is the EPI contribution at 0 K at the appropriate volume \cite{Varma2022, Allen2022EPIBS}, the present paper allows the accurate computation of the free-energy at finite temperatures, by including the EPI contributions. 
    
    Thus, the present work opens the way to compute the most accurate total energy, at 0 K, and free-energies at finite temperatures. Hence, it will be of great importance in accurately determining the energy differences between polymorphs, phase transition temperatures and pressures etc. for the applications discussed above \cite{Oganov2019structure, 	Catlow2020structure, Cheng2022crystal, Hong2020training, Yamashita2018crystal, Kawanishi2016, Scalise2019, Ramakers, Cazorla2019, 	Van2007, Bhat2018}.
	
	We note that a new formalism for the \textit{ab initio} total energy, including EPI contributions, has been recently proposed \cite{Ponce2025EPItotal}. The origins of the standard Hamiltonian and the recent proposal are different. The standard Hamiltonian is based on the Rayleigh-Schr\"{o}dinger perturbation theory, while the other formalism is not. We note that the validity of the former, for total energy studies, has been established by Allen and Hui \cite{Allen1980}. A comparison of the two formalisms is beyond the scope of the present paper and will be presented elsewhere. The emphasis of the present paper is on the above mentioned important aspects concerning the standard Hamiltonian.
       
	\section{Theorectical background to the EPI contribution to the total 
	energy}
	
	In the Allen-Heine theory, EPI contribution to the eigenenergy 
	arises from the Debye-Waller (DW) and Fan-Migdal (FM) terms and, at 0 K, is 
	given by  
	\cite{Allen1976, Allen1980}
	\begin{equation}
		\Delta \epsilon_{n\textbf{k}}(0) = 
		\Delta\epsilon_{n\textbf{k}}^{\textrm{DW}}(0) + 
		\Delta\epsilon_{n\textbf{k}}^{\textrm{FM}}(0)
		\label{EPI-Energy-eigenstate} 
	\end{equation}
	
	The FM self-energy term includes contributions from both conduction and 
	valence bands \cite{Allen1976, Allen1980}. It can also be written as a sum 
	of their separate contributions as \cite{Nery2018} 
	\begin{equation}
		\Delta\epsilon_{n\textbf{k}}^{\textrm{FM}}(0) = 
		\Delta\epsilon_{n\textbf{k}}^{\textrm{FM},\textrm{occ}}(0)
		+\Delta\epsilon_{n\textbf{k}}^{\textrm{FM},\textrm{unocc}}(0) 
		\label{FM-occ-unocc} 
	\end{equation}
	
	These partial-FM self-energy terms are given by \cite{Nery2018}
	\begin{equation}
		\Delta \epsilon_{n\textbf{k}}^{\textrm{FM,unocc}}(0) = \frac{1}{N_\textbf{q}} 
		\sum^{\textrm{BZ}}_{\textbf qj} \sum^{\textrm{unocc}}_{n'} 
		\frac{{|\langle\textbf{k}n|H_j^{(1)}|\textbf{k}+\textbf{q}n'\rangle|}^2}{\omega
			- \epsilon_{\textbf{k}+\textbf{q}n'}-\omega_{\textbf qj}+i\eta'}
		\label{FM-unocc}
	\end{equation}
	
	and 
	\begin{equation}
		\Delta \epsilon_{n\textbf{k}}^{\textrm{FM,occ}}(0) = \frac{1}{N_\textbf{q}} 
		\sum^{\textrm{BZ}}_{\textbf qj} \sum^{\textrm{occ}}_{n'} 
		\frac{{|\langle\textbf{k}n|H_j^{(1)}|\textbf{k}+\textbf{q}n'\rangle|}^2}{\omega
			- \epsilon_{\textbf{k}+\textbf{q}n'}+\omega_{\textbf qj}+i\eta'}
		\label{FM-occ}
	\end{equation}
	
	Because the EPI alters all the eigenenergies, we have proposed that the 
	total energy is also altered which leads to the EPI contribution to the 
	total energy \cite{Varma2022}. Subsequently, Allen has derived the formal expression for 	the EPI contribution to total energy, $\Delta E_{\textrm{EP}} (V,0)$, from second-order perturbation theory, as \cite{Allen2022erratum, Allen2023}
	\begin{equation}
    \begin{split}
		\Delta E_{\textrm{EP}} (V,0) &= 	\sum_k \Bigg[\langle k 
		|V^{(2)}|k\rangle \\&+ \sum_Q \frac{|\langle k 
			|V^{(1)}|k+Q\rangle|^2}{\epsilon_k - 
			\epsilon_{k+Q}-\hbar\omega_{-Q}}(1-f_{k+Q}) \Bigg]\;f_k
		\label{EPI-zero-K-Allen}
    \end{split}
	\end{equation}
	
	The  EPI contribution to individual eigenenergy, $\Delta 
	\epsilon_{n\textbf{k}}(0)$ (Eq. \ref{EPI-Energy-eigenstate}), summed over 
	occupied states is the EPI contribution to the band-structure energy, 
	$\Delta E^{\textrm{ep}}_{\textrm{bs}}$.	The structure of Eq. 
	\ref{EPI-zero-K-Allen} is similar, not identical, to the structure of 
	$\Delta E^{\textrm{ep}}_{\textrm{bs}}$. The first term in Eq. 
	\ref{EPI-zero-K-Allen} is the DW term. However, the second (FM) term 
	differs in the presence of the ($1-f_{k+Q}$) factor 
	\cite{Allen2022erratum}. The ($1-f_{k+Q}$) factor is present to ensure that 
	the $k+Q$ is an empty state to satisfy the Pauli principle 
	\cite{Frohlich1950theory, Fan1951}, due to which the summation is 
	only over unoccupied energy bands  \cite{Fan1951, Allen2022EPIBS}. The 
	second term can be written more explicitly as  \cite{Allen2023}
	\begin{equation}
		\Delta 
		E_{\textrm{tot}}^{\textrm{FM,unocc}} = \\2 \sum_{\vec{k}n}^{occ} 
		\sum_{\vec{k}+\vec{Q}n'}^{unocc}\sum_{j=1}^{3n_{at}}
		\frac{|M_{\vec{k}n,\vec{k}+\vec{Q}n',j}|^2}{\epsilon_{\vec{k}n} -
			\epsilon_{\vec{k}+\vec{Q}n'}-\hbar\omega_{\textbf qj}}
		\label{EPI-FM-unocc-Allen}
	\end{equation}
	where the 2 is for spin and $n_{at}$ is the number of atoms in the 
	supercell used. Eq. \ref{EPI-FM-unocc-Allen} is identical to Eq.2.11 of 
	Fr{\"o}hlich \cite{Frohlich1950theory} and Eq.4 of Fan \cite{Fan1951} in 
	the earliest EPI studies. 

	With this brief background we discuss the accurate computation of the EPI 
	contribution to the total energy.

\section{Recasting Allen's equation}	

	The  computational package used in the present work for EPI computations is the TDepES module of the  ABINIT software package \cite{Gonze2009, Gonze2016, Abinit}). This module has been widely used to obtain temperature dependent (due to EPI) band-structures and band-gaps \cite{Ponce2015, Ponce2014b, Ponce2014a, Antonius2015, Friedrich2015, Tutchton2018, Querales2019, Miglio2020}. 

    However, the TDepES module cannot compute  Eq. \ref{EPI-zero-K-Allen} due to the presence of the ($1-f_{k+Q}$) factor. We describe below a method to accurately compute  Eq. \ref{EPI-zero-K-Allen} based on the EPI implementation (TDepES) in the ABINIT software package \cite{Gonze2009, Gonze2016, Abinit}.
	
	From Eqs. \ref{EPI-Energy-eigenstate}, \ref{FM-occ-unocc} and 
	\ref{EPI-FM-unocc-Allen}, Eq. \ref{EPI-zero-K-Allen} can be rewritten as
	\begin{equation}
		\begin{split}
			\Delta E_{\textrm{tot}}^{\textrm{ep}}  &= 
			\sum^{\textrm{occ}}_{n,\textbf{k}} \left[ 
			\Delta\epsilon_{n\textbf{k}}^{\textrm{DW}} + 	
			\Delta\epsilon_{n\textbf{k}}^{\textrm{FM,unocc}}\right] \\ &=
			\sum^{\textrm{occ}}_{n,\textbf{k}} \left[\Delta 
			\epsilon_{n\textbf{k}} - 
			\Delta\epsilon_{n\textbf{k}}^{\textrm{FM,occ}}\right] \\
			&= \Delta E^{\textrm{ep}}_{\textrm{bs}} - \Delta 
			E_{\textrm{tot}}^{\textrm{FM,occ}}
			\label{EPI-zero-K-Allen-rewritten}
		\end{split}
	\end{equation}

    The band-structure energy, $\Delta E^{\textrm{ep}}_{\textrm{bs}}$, is given by
\begin{equation}
		\Delta E^{\textrm{ep}}_{\textrm{bs}} =  
		 \sum^{\textrm{occ}}_{n,\textbf{k}} \left[ 
		\Delta\epsilon_{n\textbf{k}}^{\textrm{DW}} + 	
		\Delta\epsilon_{n\textbf{k}}^{\textrm{FM}}\right]
		 =  \sum^{\textrm{occ}}_{n,\textbf{k}}  \Delta \epsilon_{n\textbf{k}}
		\label{EPI-band-struc-energy}
	\end{equation}
    
The partial-FM-occ term is given by 
\begin{equation}
\begin{split}
 \Delta E_{\textrm{tot}}^{\textrm{FM,occ}} & =  \sum_{\mathbf{k},\mathbf{q}}^{\textrm{BZ}}
\sum_{\nu}
\sum_{nm}^{occ}
\frac{\left|g_{mn\nu} (\mathbf{k}, \mathbf{q})\right|^{2}}{\varepsilon_{n\mathbf{k}} - \varepsilon_{m\mathbf{k}+\mathbf{q}} + \hbar\omega_{\textbf q\nu}} 
			\label{EPI-FM-occ-tot}
		\end{split}
	\end{equation}
where $g_{mn\nu} (\mathbf{k}, \mathbf{q})$ is the electron-phonon matrix element \cite{Giustino2017, Ponce2025EPItotal}. 

	For non-magnetic systems, each eigenstate can accommodate 2 electrons of opposite spin and hence, an additional factor of 2 is present, as explicitly represented in Eq. \ref{EPI-FM-unocc-Allen}. 
    
    The equivalent expression, Eq. \ref{EPI-zero-K-Allen-rewritten}, shows that the EPI contribution to the total energy comes from the EPI contribution to the band-structure energy, $\Delta E^{\textrm{bs}}_{\textrm{ep}}$, corrected by the partial-FM-occ contribution, $\Delta 
	E_{\textrm{tot}}^{\textrm{FM,occ}}$. The second term is effectively present due to the anti-symmetric nature of the crystal wavefunction \cite{Fan1951}.

\section{EPI contribution to the total energy in the adiabatic approximation}

	As is well known, the EPI contribution to each eigenstate can be computed in two approximations, adiabatic and non-adiabatic \cite{Ponce2015}. The main difference between the two is that in the adiabatic approximation, the phonon energy term in the denominator of Eq. \ref{FM-unocc} to Eq. \ref{EPI-zero-K-Allen} and Eq. \ref{EPI-FM-occ-tot} is neglected and set to zero. In the computation of the EPI contribution to individual eigenstates, both the adiabatic and non-adiabatic approximations can be used for infrared (IR) inactive materials, while for IR-active materials it is essential to use the non-adiabatic approximation \cite{Ponce2015}. 	

	For the EPI contribution to the total energy, Poncé and Gonze \cite{Ponce2025EPItotal} have recently shown that, in the adiabatic approximation,  the partial-FM-occ contribution is identically zero. This is because the summation over bands, $m$ and $n$, are only over occupied bands, and upon exchange of $n\textbf{k}$ and $m\textbf{k+q}$, there will be a change in sign of the denominator, $\epsilon_{n\textbf{k}} - \epsilon_{m\textbf{k+q}}$. Hence, the  $\Delta 
E_{\textrm{tot,adia}}^{\textrm{FM,occ}}$ is identically zero \cite{Ponce2025EPItotal}. 

Thus, in the adiabatic approximation, Eq. \ref{EPI-zero-K-Allen-rewritten} becomes
\begin{equation}
			\Delta E_{\textrm{tot}}^{\textrm{ep,adia}}  = 
			 \Delta E^{\textrm{ep}}_{\textrm{bs,adia}} 
			\label{EPI-tot-adia}
	\end{equation}
That is, in the adiabatic approximation, the  EPI contribution to the band-structure energy, $\Delta E^{\textrm{ep}}_{\textrm{bs,adia}}$, is the sole term in the EPI contribution to the total energy, $\Delta E_{\textrm{tot}}^{\textrm{ep}}$. This was first proposed by Fan [33], and also independently by us in our earlier work [18].

However, Allen \cite{Allen2022EPIBS} has shown 	that if the EPI contribution to the band-structure energy, $\Delta E^{\textrm{ep}}_{\textrm{bs}}$, is the sole term in the  EPI contribution to the total energy, $\Delta E_{\textrm{tot}}^{\textrm{ep}}$, it is theoretically invalid as it can be derived from the free-energy expression for non-interacting quasiparticles.
	
The following derivation is reproduced from Allen \cite{Allen2022EPIBS}. 
The free energy shift caused by  the shift in eigenenergies due to the EPI, $\Delta \epsilon _K^{\textrm{ep}}$, is,

\begin{equation}
\begin{split}
	\Delta F_{\text{epi}} =& k_B T\sum_{K}  \ln\left[\frac{1 - 
			f(\epsilon_K^{QH} + \Delta \epsilon_K^{\textrm{ep}})}{1 - 
			f(\epsilon_K^{QH})}\right] \\      
    \approx & \sum_{K} f_K^{QH} \Delta 
	\epsilon_K^{\textrm{ep}} \rightarrow \Delta 
		E^{\textrm{ep}}_{\textrm{bs}}
	    \label{EPI-non-int-1}
        \end{split}
        \end{equation}
 	when expanded to the lowest order \cite{Allen2022EPIBS}. 
 
Another way to arrive at the same conclusion is described below. From the 
free energy expression for non-interacting electrons, it is readily 
seen that at 0 K, the (free) energy is 
	\begin{equation}
		E^{\textrm{non-int}}_{\textrm{el}} = \sum_{K} (\epsilon_K-\mu) 
		f({\epsilon_K}) 
		\rightarrow E_{\textrm{el}}^{\textrm{DFT}}
	\label{EPI-non-int-2}
	\end{equation}
This is also the expression for the total energy of non-interacting electrons at 0 K given in standard books \cite{Ashcrof1976, Simon2013oxford, Martin2020electronic, Grosso2013solid}. The band-structure 
energy in Eq. \ref{EPI-non-int-2} is replaced in Eq.11 of Allen 
\cite{Allen2020} by the more accurate $E_{\textrm{el}}^{DFT}$. When the EPI contribution to each eigenstate is included, the electron eigenenergy  
$\epsilon_K$ is substituted by $\epsilon_K+\Delta 
 \epsilon_K^{\textrm{ep}}$. Then, the total energy of electrons, in 
 semiconductors and insulators, is given by
	\begin{equation}
	\begin{split}
		E^{\textrm{non-int}}_{\textrm{el}} &= \sum_{K} (\epsilon_K + \Delta 
	\epsilon_K^{\textrm{ep}} -\mu) f({\epsilon_K}) \\ & \approx \sum_{K} 
	(\epsilon_K 
	-\mu) f({\epsilon_K}) + \sum_{K} \Delta 
	\epsilon_K^{\textrm{ep}} f({\epsilon_K}) \\
		&\rightarrow E_{el}^{\textrm{DFT}} + \sum_{K} \Delta 
	\epsilon_K^{ep} f({\epsilon_K}) \\
	&\rightarrow E_{\textrm{el}}^{\textrm{DFT}} + \Delta 
	E^{\textrm{\textrm{ep}}}_{\textrm{bs}}
	\label{EPI-non-int-3}
	\end{split}
	\end{equation}

	Thus, it is seen, from the above methods, that if $\Delta E^{\textrm{ep}}_{\textrm{bs}}$ is the sole EPI contribution 
	to 	the total energy,  $\Delta E_{\textrm{tot}}^{\textrm{ep}}$, the result can be derived ``using the formula for free energy of independent electrons evaluated with the quasiparticle energy” \cite{Allen2022EPIBS}. Thus, within second-order perturbation theory, the EPI contribution to the total energy in the adiabatic approximation, Eq. \ref{EPI-tot-adia}, is theoretically invalid \cite{Allen2022EPIBS} for all materials, IR-active or IR-inactive. 

    Recently, a fourth-order contribution to the EPI energy, $E^{\mathrm{elph}}$ has been derived \cite{Ponce2025EPItotal}. The $E^{\mathrm{elph}}$ term contains a summation only over unoccupied states and cannot be obtained from the free energy expression for non-interacting particles. Therefore, for the EPI contribution to the total energy in the adiabatic approximation to be theoretically valid, it is necessary to go beyond second-order perturbation theory (and the standard Hamiltonian) and include fourth-order terms. 

\section{EPI contribution to the total energy in the non-adiabatic 
approximation}

In the non-adiabatic approximation, the phonon energy term has to be included. In this case, the partial-FM-occupied contribution to the total energy of the crystal, $\Delta E_{\textrm{tot,non-adia}}^{\textrm{FM,occ}}$, is given by Eq. \ref{EPI-FM-occ-tot} (including the phonon energy term).

Similar to the discussion in the previous section, upon exchange of 
$n\textbf{k}$ and $m\textbf{k+q}$, there will be a change in sign for 
$\epsilon_{n\textbf{k}} - \epsilon_{m\textbf{k+q}}$ in the denominator. 
However, the presence of the phonon energy term in the denominator, which remains unchanged in the above process, ensures that complete cancellation does not occur. Hence, $\Delta E_{\textrm{tot,non-adia}}^{\textrm{FM,occ}}$ is a finite quantity. It also cannot be derived from the free energy expression for non-interacting particles. Therefore, the EPI contribution to the total energy, $\Delta E_{\textrm{tot}}^{\textrm{ep}}$, Eq. \ref{EPI-zero-K-Allen-rewritten}, is theoretically valid in the non-adiabatic approximation as it addresses Allen's objections discussed above.

Given that $\Delta E_{\textrm{tot,adia}}^{\textrm{FM,occ}}$ is identically 
zero, $\Delta E_{\textrm{tot,non-adia}}^{\textrm{FM,occ}}$ is likely to be a small quantity. This can be explored further by rewriting Eq. \ref{EPI-FM-occ-tot} so that the the two terms that change signs are paired together as follows
\begin{equation}
\begin{split}
  \Delta E_{\textrm{tot,non-adia}}^{\textrm{FM,occ}} = \frac{1}{2}
  \sum_{\mathbf{k},\mathbf{q}}^{\textrm{BZ}}
&\sum_{\nu}
\sum_{nm}^{occ}
\Bigg[\frac{\left| g_{mn\nu}(\mathbf{k}, \mathbf{q}) 
\right|^{2}}{\varepsilon_{n\mathbf{k}} - \varepsilon_{m\mathbf{k}+\mathbf{q}} 
+ \hbar\omega_{\mathbf{q}\nu}}  \\
&+ \frac{\left| g_{mn\nu}(\mathbf{k}, \mathbf{q}) 
\right|^{2}}{\varepsilon_{m\mathbf{k}+\mathbf{q}} - \varepsilon_{n\mathbf{k}} 
+ \hbar\omega_{\mathbf{q}\nu}}\Bigg]
\label{FM-occ-Non-Adia}
\end{split}
\end{equation}
It is readily seen that upon neglecting the the phonon energy term in the 
denominator, Eq. \ref{FM-occ-Non-Adia} becomes  $\Delta 
E_{\textrm{tot,adia}}^{\textrm{FM,occ}}$ and is identically zero, as required.

By taking a common denominator and with algebraic manipulation, Eq. 
\ref{FM-occ-Non-Adia} can be written as 
\begin{equation}
    \Delta E_{\textrm{tot,non-adia}}^{\textrm{FM,occ}} = - \frac{1}{2}
\sum_{\mathbf{k},\mathbf{q}}^{BZ}
\sum_{\nu}
\sum_{nm}^{\mathrm{occ}}
\frac{2 \, \hbar\omega_{\nu \mathbf{q}} \,
\left| g_{mn\nu}(\mathbf{k}, \mathbf{q}) \right|^{2}}
{\left(\varepsilon_{n\mathbf{k}} - \varepsilon_{m\mathbf{k}+\mathbf{q}} 
\right)^{2} - \left(\hbar\omega_{\textbf q\nu}\right)^2} 
\label{FM-occ-NAdia-4th-order}
\end{equation}
\\
Comparing the two equivalent expressions, Eq. \ref{EPI-FM-occ-tot} and Eq. 
\ref{FM-occ-NAdia-4th-order}, we see Eq. \ref{FM-occ-NAdia-4th-order} is to be 
preferred as it incorporates complete physics underlying the partial-FM-occ 
term.

The structure of Eq. \ref{FM-occ-NAdia-4th-order} is very similar to that of the fourth-order term, $E^{\mathrm{elph}}$ \cite{Ponce2025EPItotal},  and differs mainly in the presence of the phonon energy term in the denominator, which is present as the former is in the non-adiabatic approximation.

Eq. \ref{FM-occ-NAdia-4th-order} shows that the net result of the incomplete cancellation of two second-order terms (in Eq. \ref{FM-occ-Non-Adia}) need not be a second-order term. Rather, it has the structure of a higher-order term and hence, its contribution to the EPI total energy is likely to be small. Therefore, in second-order perturbation theory, that is the basis of the standard Hamiltonian, the $\Delta E_{\textrm{tot,non-adia}}^{\textrm{FM,occ}}$ term can be neglected. In this regard, we point out that Allen and Hui \cite{Allen1980} have neglected the first term in Eq.11 of Ashkenazi et al. \cite{Ashkenazi1979}, which has $energy^2$ in the denominator and the numerator is similar (not identical) to Eq. \ref{FM-occ-NAdia-4th-order} above, stating that it is a higher order term that can be neglected in second-order perturbation theory.

From a computational perspective, due to cancellation of errors, the neglect of the $\Delta E_{\textrm{tot,non-adia}}^{\textrm{FM,occ}}$ term, which is a higher-order term, is likely to contribute marginally to the error in energy differences between polymorphs. Therefore, this term is likely to be of importance only when the energy differences between polymorphs are marginal. In such cases,  the $\Delta E_{\textrm{tot,non-adia}}^{\textrm{FM,occ}}$ term must also be considered and computed.

It follows from Eq. \ref{EPI-zero-K-Allen-rewritten} that the EPI contribution to the band-structure energy,  $\Delta E^{\textrm{ep}}_{\textrm{bs,non-adia}}$, becomes the sole EPI contribution to the total energy, $\Delta E_{\textrm{tot}}^{\textrm{ep,non-adia}}$, in the non-adiabatic approximation, that is correct to second-order. 

\section{Comparison of the two approximations for total energy studies}

From the above discussions, it is clear that, in total energy studies, the adiabatic approximation is theoretically invalid, for all materials, within second-order perturbation theory (or the standard Hamiltonian), due to Allen's objections. 

In contrast, the non-adiabatic approximation is theoretically valid within the standard Hamiltonian as it addresses Allen's objections. In this approximation, the partial-FM-occ term can be rewritten which shows that it has the structure of a higher order term. Because the standard Hamiltonian is based on second-order perturbation theory, this term can be neglected. This is especially the case for determining the energy differences between polymorphs, as discussed above. This approximation leads to the EPI contribution to the band-structure energy as the sole EPI contribution to the total energy and does not address Allen's objections. However, this invalidity is due to computational approximations and not due to theoretical reasons.

\textit{Therefore, in total energy studies based on the standard Hamiltonian,  the non-adiabatic approximation must be preferred for all materials, IR-active and IR-inactive, on theoretical grounds.} 

\section{Theoretical aspects underlying computation of EPI 
contribution to total energy}

In the computation of the EPI contribution to each eigenenergy, a small 
imaginary term $i\delta$  (of 10 - 100 meV) is added to smoothen the energy 
denominators and improve convergence \cite{Ponce2015, Cardona992isotope, Giustino2010, Antonius2014, Abinit}. This parameter can be varied and should be small to obtain accurate values \cite{Ponce2015}. 

Thus, the partial-FM-occ term in the adiabatic approximation is to be computed from the expression 
\begin{equation}
\begin{split}
  \Delta E_{\textrm{tot,adia}}^{\textrm{FM,occ,comp}} = 
  \sum_{\mathbf{k},\mathbf{q}}^{\textrm{BZ}}
\sum_{\nu}
\sum_{nm}^{occ}
\frac{\left|g_{mn\nu} (\mathbf{k}, \mathbf{q})\right|^{2}}{\varepsilon_{n\mathbf{k}} - \varepsilon_{m\mathbf{k}+\mathbf{q}}  + (i\delta)}  
\label{Ponce-Comp-Eq 111}
\end{split}
\end{equation}

It is readily seen that, because of the $i\delta$ term, the cancellation is not exact and $\Delta E_{\textrm{tot,adia}}^{\textrm{FM,occ}}$ is likely to be non-zero. Following the above discussion of Eq. \ref{FM-occ-Non-Adia} and Eq. \ref{FM-occ-NAdia-4th-order}, the above equation, Eq. \ref{Ponce-Comp-Eq 111}, can be rewritten by combining the two terms where the eigenenergies in the denominator change signs upon exchange of $n\textbf{k}$ as $m\textbf{k+q}$ as
\begin{equation}
    \Delta E_{\textrm{tot,adia}}^{\textrm{FM,occ,comp}} = - \frac{1}{2}
\sum_{\mathbf{k},\mathbf{q}}^{BZ}
\sum_{\nu}
\sum_{nm}^{\mathrm{occ}}
\frac{2 \, (i\delta) \,
\left| g_{mn\nu}(\mathbf{k}, \mathbf{q}) \right|^{2}}
{\left(\varepsilon_{n\mathbf{k}} - \varepsilon_{m\mathbf{k}+\mathbf{q}} 
\right)^{2} - (i\delta)^2} 
\label{FM-occ-Adia-comp-4th-order}
\end{equation}

As a check, both Eq. \ref{Ponce-Comp-Eq 111} and Eq. \ref{FM-occ-Adia-comp-4th-order} both go to zero when the smearing parameter is set to zero. We see that Eq. \ref{FM-occ-Adia-comp-4th-order} has the structure of a higher-order term. While both Eq. \ref{Ponce-Comp-Eq 111} and Eq. \ref{FM-occ-Adia-comp-4th-order} are the same, Eq. \ref{FM-occ-Adia-comp-4th-order} is to be preferred for computations. This is because in Eq. \ref{FM-occ-Adia-comp-4th-order}  each term will be relatively smaller than in Eq. \ref{Ponce-Comp-Eq 111} because it is the remainder of incomplete cancellation of two terms in Eq. \ref{Ponce-Comp-Eq 111}.  Because the theoretical value is known to be zero (or small), Eq. \ref{FM-occ-Adia-comp-4th-order} is to be preferred for computation of $ \Delta E_{\textrm{tot,adia}}^{\textrm{FM,occ,comp}}$.

Using similar arguments, the partial-FM-occ term in the non-adiabatic 
approximation is altered by including the smearing parameter to 
\begin{equation}
\begin{split}
    &\Delta E_{\textrm{tot,non-adia}}^{\textrm{FM,occ,comp}} \\ &= - \frac{1}{2}
\sum_{\mathbf{k},\mathbf{q}}^{BZ}
\sum_{\nu}
\sum_{nm}^{\mathrm{occ}}
\frac{2 \, (\hbar\omega_{\nu \mathbf{q}} + i\delta) \,
\left| g_{mn\nu}(\mathbf{k}, \mathbf{q}) \right|^{2}}
{\left(\varepsilon_{n\mathbf{k}} - \varepsilon_{m\mathbf{k}+\mathbf{q}} 
\right)^{2} - \left(\hbar\omega_{\textbf q\nu} + (i\delta)\right)^2} 
\label{FM-occ-NA-comp-4th-order}
\end{split}
\end{equation}

Using similar arguments, it is seen that that new expression, Eq. 
\ref{FM-occ-NA-comp-4th-order}, incorporates the complete physics underlying 
this term and also contains relatively smaller terms than the standard 
expression (Eq. \ref{EPI-FM-occ-tot} with smearing parameter $(i\delta)$ term 
added in the denominator) and hence, it is to be preferred for computational 
purposes.

Further, because Eq. \ref{FM-occ-Adia-comp-4th-order} tends to zero as the smearing parameter $(i\delta)$ tends to zero, it should be used to determine the value of smearing parameter, $(i\delta)$ to be used in Eq. \ref{FM-occ-NA-comp-4th-order}. This will allow the effect of the phonon energy  to be separated from that of the smearing parameter.

\section{Computational details}	
		
	We report the results of the computations performed for the carbon polytypes, C-dia and C-hex (Lonsdaleite). All the computations were performed using the ABINIT software package \cite{Gonze2009, Gonze2016, Abinit} with the version, v.10.2.3. The TDepES module was used to compute the EPI contributions in the non-adiabatic approximation. The ONCV \cite{ONCVHamann,ONCVPseudodojo} pseudopotential with PBE \cite{PerdewPBE} exchange-correlation functional was used.
    
    Unshifted grids 8x8x8 and 9x9x5 were used for the C-dia and C-hex respectively. An energy cutoff of 50 Hartree was used for both structures. The q-point grid used were 30x30x30 and 31x31x19 for the C-dia and C-hex respectively. The optimized number of bands used for EPI computations was 22 for both C-dia and C-hex structures.  
	
	\section{Results and Discussion}
	The C-dia and C-hex structures are analogous to the zincblende and wurtzite structures respectively and vary only in the orientation of the $sp^3$-bonded tetrahedra \cite{Zunger1992zinc, Bechstedt2002properties, Bechstedt1994electronic}. For this reason, the C-dia primitive unit cell contains two Carbon atoms and the  C-hex primitive unit cell contains four Carbon atoms. Given their structural similarities, it follows that the energy differences for the 	various energy terms are likely to be small.
    
    The lattice parameter for C-dia structure is 6.75 Bohr. The lattice parameters for the C-hex structure are 4.75 and 7.90 Bohr. The carbon atoms occupy the 4f Wyckoff position with z = 0.06273 of space group 194, P$6_3/mmc$. These values are similar to values reported in literature \cite{Varma2022}.
    
     In Table \ref{table1} we report the EPI contribution to the band-structure energ, $\Delta E^{\textrm{ep}}_{\textrm{bs}}$, in the non-adiabatic approximation for various smearing parameters $(i\delta)$. 
     
    \begin{table}[h]
		\caption{\label{table1}%
			The EPI  contribution to the band-structure energy $\Delta E^{\textrm{ep}}_{\textrm{bs}}$ in the non-adiabatic approximation for C-dia and C-hex primitive cells and their difference using different smearing parameters, $(i\delta)$.  All energies are in meV/atom.}		
		\begin{tabular}{c c c c}
			\hline\hline
			$(i\delta)$ &\textbf{C-dia}  &\textbf{C-hex} & $\Delta (\Delta  E^{\textrm{ep}}_{bs})$ \\ 
			\hline
			  5 meV  & -11.5& -5.7 & -5.8\\
            10 meV  & -14.6 & -4.2 & -10.4\\
            100 meV  & -28.6 & -5.3 & -23.3\\
            \hline\hline
		\end{tabular}
    \end{table}

    The results reported in  Table \ref{table1} allow the computation of the standard Hamiltonian. \textit{In Table \ref{table2} we report, for the first time, the computation	of the standard Hamiltonian of a coupled-EP system, at 0 K.} The various terms in the standard Hamiltonian, Eq. \ref{Std-Hamiltonian-EPI}, are reported below.

	\begin{table}[h]
		\caption{\label{table2}%
			The various contributions to the standard Hamiltonian of a 
			coupled-EP system, Eq. \ref{Std-Hamiltonian-EPI}, at 0 K, for C-dia and C-hex polytypes. The EPI term is taken from Table \ref{table1}. The relative stability due to each energy term is also reported. All energies are in eV/atom.}		
		\begin{tabular}{c c c c c}
			\hline\hline
			Structure & $E_{DFT}$  & $E_{ZPVE}$ &$\Delta E ^{\textrm{ep,non-adia}}_{\textrm{bs}}$ & $E_{TOT}$ \\
			\hline
			C-dia 
			& -163.780& 0.179 & -0.015 &  -163.630\\ \hline
			C-hex
			& -163.756 & 0.178 & -0.004 & -163.583 \\ \hline
			\textbf{Stability}
			& -0.024 & 0.001 & -0.011  & \textbf{-0.034} \\ \hline
			\hline\hline
		\end{tabular}
    \end{table}

	The DFT relative stability (24 meV/atom) of C-dia is similar to reported values \cite{Needs2015low}. The ZPVE values are consistent with literature values \cite{Scheffler2018performance, Grochala2014diamond}. The EPI contribution to the phonon energy is not included in the standard Hamiltonian, Eq. \ref{Std-Hamiltonian-EPI}, because it is already included explicitly in DFPT and implicitly in FDM and other methods \cite{Allen2020, Allen2022EPIBS, Heid201312, Giustino2017, Varma2022}. 

The EPI contribution, $\Delta E^{\textrm{ep,non-adia}}_{\textrm{bs}}$, makes the C-dia structure more stable by 11 meV/atom. Table \ref{table1} also shows that the EPI term contributes more than ZPVE to energy differences and must be included in all \textit{ab initio} total energy studies. 

The final total energy of the crystal computed using the standard Hamiltonian shows that the C-dia structure is more stable than the C-hex structure by 34 meV/atom.

As seen from the discussion of Eq. \ref{FM-occ-NAdia-4th-order}, the neglected term, $\Delta E_{\textrm{tot,non-adia}}^{\textrm{FM,occ}}$ has the structure of a higher order term. Further, for determining the relative stability of polymorphs, the energy differences are required. This implies the energy difference between two higher than second order terms. Thus, the neglected term, $\Delta E_{\textrm{tot,non-adia}}^{\textrm{FM,occ}}$ is likely to contribute marginally to energy differences.

Thus, the total energy difference and the relative stability reported in Table \ref{table1} are accurate to second-order. (As discussed earlier, the neglected term, $\Delta E_{\textrm{tot,non-adia}}^{\textrm{FM,occ}}$, must be computed only when the energy difference between polymorphs is marginal.)

As discussed elsewhere \cite{Varma2022}, to include the EPI contributions at finite temperatures for large band-gap semiconductors and insulators, it is only necessary to add the EPI contribution to the total energy at 0 K at the appropriate volume to the the Quasi-Harmonic Approximation (QHA) \cite{Nath2016, Togo2015, Allen2020} expression. Thus, the present work, that demonstrates the computation of the EPI contribution to the total energy, $\Delta E^{\textrm{ep,non-adia}}_{\textrm{tot}}$, will lead to more accurate free energy computations at finite temperatures than the QHA.
\\

	\section{Conclusion}
	In conclusion, for the first time, we report the computation of the 
	standard Hamiltonian of a coupled EP system. For this purpose, we compute 
	accurately, to second-order, the EPI contribution to the total energy. This 
	gives the most accurate \textit{ab initio} total energy till date. We show 
	that, unlike for the EPI contribution to individual eigenstates, for the 
	EPI contribution to the total energy, the non-adiabatic approximation must 
	be used for all materials, IR-active and IR-inactive. We have derived new 
	expressions for the partial-FM-occ term, for both theoretical and 
	computational studies. The new theoretical expression shows that the 
	partial-FM-occ term has the structure of a higher-order term and not a 
	second-order term. Thus, it can be neglected unless the energy differences 
	between polymorphs is marginal. In theoretical and computational studies, 
	the new expressions for the partial-FM-occ term must be preferred over the 
	standard expressions  because they incorporate the complete physics 
	underlying this term. The standard Hamiltonian computed for the C-dia and 
	C-hex polymorphs shows that the C-dia polytype is more stable than the 
	C-hex polytype by 34 meV/atom. Our results are accurate to second-order. 
	The present work also opens the way to compute more accurate free-energy at 
	higher temperature, because only the EPI contribution at 0 K, at the 
	appropriate volume, needs to be added to the free-energy obtained in the 
	Quasi-Harmonic Approximation.
	\\
	
	We thank Prof. P. B. Allen for sharing a corrected version of Eq. 
	\ref{EPI-zero-K-Allen}. We thank Prof. Samuel  Poncé  for sharing Ref. 
	\cite{Ponce2025EPItotal}.   We thank Prof. Sumiran Pujari for helpful 
	discussions. We thank the ``Paramrudra'' HPC facility of IIT Bombay for 
	computational support.

\bibliography{references.bib}

\end{document}